# Zeroth order quantum coherences and preparation of pseudopure state in homonuclear dipolar coupling spin systems

G.B. Furman

*Department of Physics, Ben Gurion University, Beer Sheva 84105, Israel*
Email:gregoryf@bgu.ac.il

Dynamics of zeroth order quantum coherences and preparation of the pseudopure states in homonuclear systems of dipolar coupling spins is closely examined. It has been shown an extreme important role of the non-diagonal part of zeroth order coherence in construction of the pseudopure state. Simulations of the preparation process of pseudopure states with the real molecular structures (a rectangular ( 1 -chloro- 4 - nitrobenzene molecule ), a chain (hydroxyapatite molecule), a ring (benzene molecule), and a double ring (cyclopentane molecule)) open the way to experimental testing of the obtained results.

multiple-quantum dynamics, pseudopure state

1. **Introduction**

Coherent control of multiple coupled quantum systems is expected to lead to profound insights in physics as well as to novel applications, such as quantum computation, quantum communication, and quantum amplification. Such coherent control is a essential goal in various fields of quantum physics, but many of the advanced theoretical and experimental results are coming from one of the oldest areas of quantum physics: nuclear magnetic resonance ( $NMR$ ). One of such result is a so-called pseudopure state was originally proposed for nuclear magnetic resonance quantum computing [1, 2].

Conveniently, the quantum algorithms start with a pure ground state where populations of all states except the ground state are equal zero. In $NMR$ however because of the small energy gaps, it is not possible to realize a pure state, where in the whole population is in one energy level, since it requires very low temperatures as well as very high magnetic fields. However, an alternate solution was discovered to overcome this problem is a so-called pseudopure state [1, 2]. The density matrix of the spin system in this state can be partitioned into the two parts. One part of the matrix is a scaled unity one, but the second part corresponds to a pure state. Such pseudopure state imitates the pure ground states. The scaled unit matrix does not contribute to observables and is not changed by unitary evolution transformations. Therefore, the behavior of a system in a pseudopure state is exactly the same as it would be in the pure state.

Recently, an elegant method of creating pseudopure states starting from an equilibrium state was proposed, which does not require a resolved equilibrium spectrum [3]. It is based on multiple-quantum ($MQ$) dynamics with filtering of the highest-order multiple-quantum coherence. The highest-order multiple-quantum coherence ( $HOMQC$ ) is constructed of just two states, $|\uparrow\rangle$ and $|\downarrow\rangle$ , where $|\uparrow\rangle$ is the state with all spins up and $|\downarrow\rangle$ is the state with all spins down. This method has been used to prepare a pseudopure state in a of six-ring homonuclear cluster, and of seven- and twelve-heteronuclear dipolar-coupled spins: single- and fully- labeled $^{13}C$ -benzene in liquid-crystalline solvent [4, 5]. However, using the double-quantum effective Hamiltonian [6, 7] the $HOMQC$ can be excited only in clusters of $2 + 4n$ coupled indentical spins, ( $n = 0, 1, 2, 3...$ ) [7, 8]. Furthermore, as we will show, a component critical to the operation of the method is reducing to zero the intensity of non-diagonal part of zero quantum ( $0Q$ ) coherence.

In the present paper the technique of preparation of the pseudopure states by using $MQ$ $NMR$ will be extended to clusters of various even numbers of coupling homonuclear spins. A comprehensive study of the pseudopure state preparation process and results of simulations for one dimensional ($1D$ ) spin clusters with up to ten spins are presented.

2. **Dynamics of zeroth-quantum coherences**

We shall consider the MQ dynamics of spin-half ( $I = 1/2$ ) system in an external magnetic field $\vec{H}_0$ with the Hamiltonian

$$H = H_0 + H_d, \tag{1}$$

where $H_0$ is the Zeeman interaction Hamiltonian,

$$H_0 = \omega_0 I_z, \tag{2}$$

$\omega_0 = \gamma |\vec{H}_0|$ is the Larmor frequency $(y = 1)$, $\gamma$ is the gyromagnetic ratio, $I_j^z$ is the angular momentum operators in the $Z$ a direction. The secular part of the dipolar Hamiltonian in a high magnetic field for indentical nuclei is

$$H_{dd} = \sum_{j<k} D_{jk} \left[ I_j^z I_k^z - \frac{1}{4} \left( I_j^+ I_k^- + I_j^- I_k^+ \right) \right], \tag{3}$$

where $D_{jk} = \frac{\gamma^2 \gamma}{2 r_{jk}^3} \left( 1 - 3\cos^2 \theta_{jk} \right)$ is the coupling constant between spins $j$ and $k$, $r_{jk}$ is the distance between spins $j$ and $k$, $\theta_{jk}$ is the angle between the internuclear vector $\vec{r}_{jk}$ and the external magnetic field $\vec{H}_0$. $I_j^+$ and $I_j^-$ are the raising and lowering spin angular momentum operators of spin $j$.

The typical scheme of the *MQNMR* experiment is the following: a spin system at equilibrium state in a strong constant magnetic field is acted upon in a time $t$ by the multiple-pulse sequence with an eight-pulse cycle [6, 7], called preparatory period. The system undergoes next, for a time $t_1$, a free evolution is driven by dipole-dipole interactions ( *DDI* ). Because the *MQ* coherences do not generate a magnetization, they are not directly measurable. Hence a detecting multiplepulse sequence is then applied, followed by the action of a $90^0$ pulse, and the transverse magnetization is measured. In a rotating frame the average Hamiltonian describing the multiple-quantum dynamics at the preparatory period can be written in the form [7]:

$$H_{av} = -\frac{1}{2} \sum_{j<k} D_{jk} \left( I_j^+ I_k^+ + I_j^- I_k^- \right) \tag{4}$$

The effect of the sequence of irradiating pulses on the spin system can be represented by the following equation

$$\rho(t) = e^{-iH_{av}t} \rho(0) e^{iH_{av}t}, \tag{5}$$

where $\rho(0)$ is the initial density matrix of the spin system. The experimentally observed values are the intensities of multiple-quantum coherences $J_{kQ}(t)$:

$$J_{kQ}(t) = \frac{1}{Tr I_z^2} \sum_{p,q} \rho_{p,q}^2(t) \quad \text{for } k = m_{zp} - m_{zq}, \tag{6}$$

where $m_{zp}$ and $m_{zq}$ are the eigenvalues of the $I_z$. Besides of the typical initial equilibrium conditions usually used in the *MQ NMR* experiments, when the spin system is described in a high temperature approximation by a density matrix $\rho_{eq}(0) = I_z$ [6, 7], we shall consider also a mixture of two states

$$\rho_{int}(0) = \pm |\uparrow\rangle\langle\uparrow| \sim |\downarrow\rangle\langle\downarrow| \qquad (7)$$

as an initial condition [9]. The state $\rho_{int}(0)$ can be considered as an intermediate state. To transform the intermediate state (7) to the pseudopure one a non-unitary operation, such as a partial saturation, is needed, which redistributes, for example, the over population of the state $|\uparrow\rangle\langle\uparrow|$ among other states but does not change the population of the state $|\downarrow\rangle\langle\downarrow|$ [3].

In the spin-1/2 system, the equilibrium density matrix for an ensemble soaking in the strong magnetic field, $\ddot{H}_0$, has only diagonal elements. The presence of these elements is said to represent zeroth quantum coherence ( $0Q$ ) [6, 7]. Zeroth-order quantum coherence can be defined according the equation (6) with $k = 0$. In the general case, not only diagonal element of the density matrix $\left(\rho_{p,p}(t)\right)$ satisfy this equation with $k = 0$, but also non-diagonal elements $\left(\rho_{p,q}(t) \text{ with } p \neq q\right)$ have the same properties [10]. For our consideration it is very convenient to divide contributions for $0Q$ intensity which comes from diagonal ( $D0Q$ ) ( $J_{0Q_{diag}}$ ) and non-diagonal ( $ND0Q$ )( $J_{0Q_{non-diag}}$ ) of the matrix elements of the density matrix at time $t$. On the one hand, both coherences, $D0Q$ and $ND0Q$, have the same symmetry properties but their dynamics is very different.

To show the difference in the evolution of the intensities of $D0Q$ -coherence ( $J_{0Q_{diag}}$ ) and of $ND0Q$ -coherence ( $J_{0Q_{non-diag}}$ ) we shall consider rings of nuclear spins in an external magnetic field $\ddot{H}_0$ perpendicular to the plane of the ring. We examine the evolution of $0Q$ -coherences by letting the spin system evolve under a Hamiltonian (4) with the dipolar coupling constant $D_{jk} = D_1 \left[\frac{\sin\frac{\pi}{N}}{\sin\left(\frac{\pi}{N}(k-j)\right)}\right]^3$, here $D_1$ is the coupling strength between the nearest spins. After calculating the evolution (Eq. (5)), we grouped the matrix elements in the density matrix $\rho_{p,q}(t)$ according Eq. (6) with $k = 0$. The results of numerical simulations are presented in Fig. 1. For $t = 0$ the diagonal part determines completely the $0Q$ spectral intensity (Fig. 1a). Then, in the initial period of evolution, $0Q$ coherences arising from non-diagonal elements of the spin density operator appear (Fig. 1b). Despite of very different time evolution of the $J_{0Q_{diag}}$ – and $J_{0Q_{non-diag}}$ – intensities, it is impossible to separate off between them using standard filtering technique [6, 7] which is based on the symmetry properties. This leaves us with only one possibility to separate the non-diagonal part of the $0Q$ -coherences from the diagonal one, the evolution of the spin system under the influence of the averaged Hamiltonian (4). During the evolution, the intensity of non-diagonal zeroth-order coherences, ( $J_{0Q_{non-diag}}$ ), may reach its zero value. For example, in the four spin ring the intensity of non-diagonal part of the $0Q$ coherence reaches its zero value at the times $t = 3.27$, $6.03$, and $9.37$ (Fig.1b). One could anticipate that there exist a strong relation between the zeroth value of $J_{0Q_{non-diag}}$ intensities and preparation the pseudopure state of a spin system.

## 3. Preparation of the pseudopure states

Detailed testing of preparation pseudopure state let us start with simulation of the $MQ$ dynamics in the four spin cluster, which can be represented by spins of 4 hydrogen nuclei of 1 -chloro- 4 -nitrobenzene molecule with the dipole-dipole coupling constants are given by $D_1 = D_{12} = D_{34} = 1$, $D_{13} = D_{24} = 1/8$, and $D_{14} = D_{23} = \frac{1}{3\sqrt{3}}$. Figure 2a shows the intensity of the non-diagonal part of $0Q$ coherence starting from the equilibrium state, $\rho_{eq}(0)$, calculated according to Eq. (5) as function of preparatory time, $\tau_1$. One can see that $J_{0Q_{non-diag}}$ reaches its zero value at times $\tau_1 = 7.86$ and $12.61$ (in units of $1/D_1$) (the arrows in figure 2a show these times). While, when the evolution starts with the initial condition (7), $\rho_{int}(0)$, the non-diagonal $0Q$ coherence does not excite, $J_{0Q_{non-diag}} = 0$ (Fig. 2b). Therefore, the strategy of preparation of a pseudopure state will consist of the following four operations: i) excitation of $MQ$ coherences starting from the equilibrium, $\rho_{eq}(0)$ and stop the preparatory period at time when the intensity of $ND0QC$ is reduced to zero (at times $\tau_1 = 7.86$ or $12.61$); ii) the $2Q$ coherence is filtering, iii) then the time reversal sequence during the times $\tau_1 = 7.86$ or $12.61$ is applied, and as final stage, iv) partial saturation is followed. Our simulation of the first three stages showed that the high mixed equilibrium state can be converted into the intermediate state (7) consists of mixture of only two states : $\left|\uparrow\right\rangle\left\langle\uparrow\right| - \left|\downarrow\right\rangle\left\langle\downarrow\right|$ with $\tau_1 = 7.86$, and $\tau_2 = 7.86$ (Fig. 3a) and $-\left|\uparrow\right\rangle\left\langle\uparrow\right| + \left|\downarrow\right\rangle\left\langle\downarrow\right|$ with $\tau_1 = 12.61$, $\tau_2 = 7.86$ (Fig. 3b).

The simulation with the spin chain were performed by using the Hamiltonian (4) with the dipolar coupling constants $D_{jk} = \frac{D_1}{(j-k)^3}$. Figure 3c shows results of the simulation of diagonal elements in a four spin chain with $\tau_1 = 84.82$, and $\tau_2 = 84.82$. The quasi-one-dimensional clusters of uniformly spaced proton spins in hydroxyapatite, $Ca_5(OH)(PO_4)_3$ [11], can be used for the experimental realization of preparation the pseudopure state in the spin chain.

The simulations show that the intermediate state (7) can not be reached when the times of the first and the third periods do not coincide with zero point of the non-diagonal part of $0Q$ - coherences. It must be stressed that the $HOMQC$ can not be exited by using the double-quantum Hamiltonian in a four spin system [8] and the pseudopure state is realized in the four spin cluster without using the $HOMQC$.

The $HOMQC$ state was used for preparation the pseudopure state in the ring of six dipolar-coupled proton spins of a benzene molecule, $C_6H_6$, dissolved in liquid crystal $ZLI1167$ [3]. As we shall see below excitation of the $HOMQC$ state is a necessary condition but insufficient one for the converting the mixed initial state to the state (7) in clusters with $2 + 4n$ coupled indentical spins. In order to confirm this point we simulate the dynamics of the $0Q$ -coherences, both arise from diagonal and non-diagonal elements of the density matrix, in the ring of six spins by using the equilibrium and mixture of two states (7) as an initial states. Figure 4 gives the time dependences of the $HOMQC$ ( $6Q$ -coherence) with equilibrium initial state (Fig. 4a) and $6Q$ - and $0Q$ - (only the non-diagonal part) coherences in the ring of six nuclear spins coupled by dipole--dipole interaction (Fig.4b). To prepare the pseudopure state the preparatory period is stopped at time $\tau_1 = 6.08$ , when the $HOMQC$ intensity reaches one of its major maxima (this time is shown by the solid arrow $a$ in figure 4b) and concurs with time when the intensity of non-diagonal $0Q$ -coherence reduces to zero. Then, all $MQ$ coherences are averaging out except the $HOMQC$ [3]. After the filtering the $HOMQC$ the time-reversal period follows with the same duration, $\tau_2 = 6.08$ , is applied. The numerical simulation results of the diagonal matrix elements of the density matrix are shown in Figure 5a and coincides with the experimental data [3]. Various combination of the time duration of preparatory, $\tau_1$ , and time-reversal , $\tau_2$ , periods were used to prepare the intermediate state (7). Results of numerical calculations are presented in Fig. 5. One can see that when the preparatory period, $\tau_1$ , corresponds to one of the major maxima of the $HOMQC$ intensity and this time coincides with the time when the $ND0QC$ is reduced to zero, the equilibrium state is practically ideally converted into the states $\pm |\uparrow\rangle\langle\uparrow| \sim |\downarrow\rangle\langle\downarrow|$ after the time-reversal period (Figs. 5a and 5b). Otherwise, when the preparatory period and time- reversal period are not corresponded to the times when the intensity of $ND0QC$ reaches its zeroth value, the conversions is far from ideal. Comparison simulations obtained with times $\tau_2 = 4.02$ (Fig. 5c) and $\tau_2 = 8.16$ (Fig. 5d) (these times are shown by the dash arrows c and d, respectively, in figure 4b) results that the deviation of diagonal matrix elements from zero (except the two elements, $\rho_{1,1}$ and $\rho_{64,64}$ ), are more than more the values of the $ND0QC$ intensities differ from zero.

This conclusion can be proved by simulation the multiple quantum dynamics for cluster consists of ten spins. It is a complex system of dipolar coupled indentical spins with 1024 quantum states. As modelling system consisting of 10 spins we shall consider spins of 10 hydrogen atoms of a cyclopentane molecule with chemical formula $C_5H_{10}$ consisting of a ring of five carbon atoms each bonded with two hydrogen atoms above and below the ring plane. We take for the numerical simulations the high-temperature limits of the dipolar coupling constants: $D_{11'} = 1$, $D_{12} = -0.178$, $D_{12'} = -0.002$, $D_{13} = -0.093$, and $D_{13'} = 0.026$ [12] . Figure 6 gives time dependences of the $10Q$ and $0Q$ (only its non-diagonal part) coherences of a cluster contains ten nuclear spins coupled by dipole-dipole interaction. To prepare the pseudopure state by using the dynamics of $HOMQC$ and $ND0QC$ intensities we will choose two characteristic times 6.59 and 9.04 (this times are shown by the dash $(t = 6.59)$ and solid $(t = 9.04)$ arrows in figure 6), when the intensity of the $HOMQC$ reaches of its maxima and the intensity of $ND0QC$ reduces to zero (at time $t = 9.04$) and does not reaches its zero value $(t = 6.59)$ (Figs. 6a and 6b). Simulations of the preparatory period with time $\tau_1 = 6.59$ where $J_{0Q_{non-diag}} \neq 0$ and the time-reversal period with the same duration does not lead to the good conversion of the initial equilibrium state to the state $\left|\uparrow\right\rangle\left\langle\uparrow\right| + \left|\downarrow\right\rangle\left\langle\downarrow\right|$ (Fig. 7a). This result connects with fact that the intensity of $ND0QC$ does not reduce to zero. At the same time, the simulations with time-reversal duration period $\tau_2 = 9.04$ where $J_{0Q_{non-diag}} = 0$ results in a very good conversion of the $\rho_{eq}(0)$ to the states $\left|\uparrow\right\rangle\left\langle\uparrow\right| - \left|\downarrow\right\rangle\left\langle\downarrow\right|$ for $\tau_1 = 6.59$ (Fig. 7b) and $-\left|\uparrow\right\rangle\left\langle\uparrow\right| + \left|\downarrow\right\rangle\left\langle\downarrow\right|$ for $\tau_1 = 9.04$ (Fig. 7c). It should be noted that the evolution of the total $0Q$ and $6Q$ coherences in a six-spin ring and $10Q$ in a ten spin cluster with the Hamiltonian of Eq. (4) and the thermal-equilibrium initial state, has been simulated previously [13, 14]. The results of our simulations with initial density matrix $\rho_{eq}(0)$ coincide with the dependences obtained in Refs. [13, 14].

## 4. Conclusion

In conclusion, we have shown that there exist a strong relation between the zeroth value of $J_{0Q_{non-diag}}$ intensities and preparation the pseudopure state of a spin system. If the duration of the time-reversal period is not concurred with the times when the intensity of $ND0QC$ reduces to zero, then the state with only two non-zero diagonal elements can not be reached. Our simulations of the preparation process of pseudopure states with the real molecular structures, such as a rectangular ( 1 -chloro- 4 -nitrobenzene molecule ), a chain (hydroxyapatite molecule), a ring (benzene molecule), and a double ring (cyclopentane molecule), confirm this conclusion and open the way to the experimental testing.

## 5. Acknowledgments

The author thanks to J.-S. Lee and A. K. Khitrin (Kent State University, Kent) for useful and stimulate discussions, and V.M. Meerovich and V.L. Sokolovsky (Ben Gurion University) for assistance in the numerical calculations. This research was supported by a grant from the US-Israel Binational Science Foundation (BSF).

**Figure captions**

Fig. 1. Time dependences (in units of $\frac{1}{D_1}$ ) of the normalized intensities of $0Q$ diagonal (a) and non-diagonal (b) parts in a rings of four (dash-dot line), six (dotted line), eight (dashed lines), and ten (solid line) nuclear spins $1/2$ coupled by DDI with the thermal equilibrium initial state.

Fig. 2. The normalized intensities of diagonal (solid line) and non-diagonal (dashed line) parts of $0Q$- coherences and $2Q$-coherence (dotted line) in a cluster of four nuclear spins $1/2$ coupled by DDI (1-chloro-4-nitrobenzene) with the initial states: a) $\rho(0) = I_z$ ; b) $\rho(0) = -e_{1,1} + e_{16,16}$. The intensity of $4Q$-coherence is equal zero.

Fig. 3. The diagonal elements of the density matrix at various combination of the time duration of the excitation $(\tau_1)$ and reversal $(\tau_2)$ periods in a cluster of four nuclear spins $1/2$ coupled by $DDI$ (1-chloro-4-nitrobenzene) with: a) $\tau_1 = 7.86$, and $\tau_2 = 7.86$ ; b) $\tau_1 = 12.61$, $\tau_2 = 7.86$ c) numerical simulation in a linear chain consisting of four spin, $\tau_1 = 84.82$, and $\tau_2 = 84.82$.

Fig. 4. The intensity of the $6Q$- coherence in a ring of six spins with the thermal-equilibrium initial state, $\rho(0) = I_z$ (a) and the intensity of the $6Q$-coherence and non-diagonal part of $0Q$-coherence with the initial state $\rho(0) = -e_{1,1} + e_{64,64}$ (b). Solid line is the $0Q$-quantum intensity which come from non-diagonal part of the density matrix ( $J_{0Qnon-diag}$ ) and dashed is the $6$-quantum intensity $J_{6Q}$.

Fig. 5. The diagonal elements of the intermediate density matrix in a ring of six spins at various combinations of the time durations of the excitation $(\tau_1)$ and reversal $(\tau_2)$ periods: a) $\tau_1 = 6.08$, $\tau_2 = 6.08$ ; b) $\tau_1 = 6.08$, $\tau_2 = 12.19$ ; c) $\tau_1 = 6.08$, $\tau_2 = 4.02$ ; c) $\tau_1 = 6.08$, $\tau_2 = 8.16$ . These times are shown by the solid (*a and b*) and dash (*c and d*) arrows in figure 4b.

Fig. 6. The normalized intensity of MQ-coherences in cyclopentane: a) non-diagonal part of $0Q$- coherences stating from the density matrix $\rho(0) = -e_{1,1} + e_{1024,1024}$ ; b), and *10*-quantum intensity with the equilibrium initial state, $\rho(0) = I_z$.

Fig. 7. The diagonal elements of the density matrix in a cyclopentane with ten spins at various combinations of the times of the preparation $(\tau_1)$ and time-reversal periods $(\tau_2)$: a) $\tau_1 = 6.59$, $\tau_2 = 6.59$ ; b) $\tau_1 = 6.59$, $\tau_2 = 9.04$ ; c) $\tau_1 = 9.04$, $\tau_2 = 9.04$ .

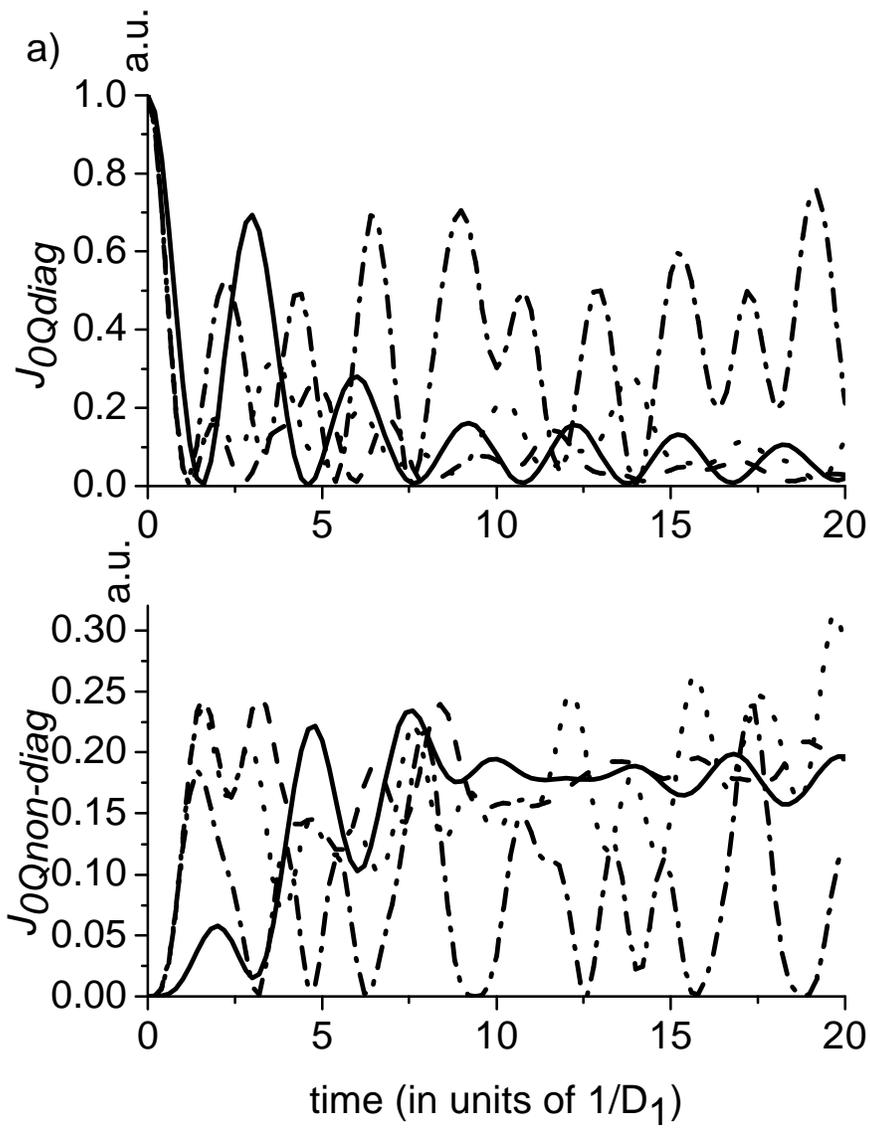

Fig.1

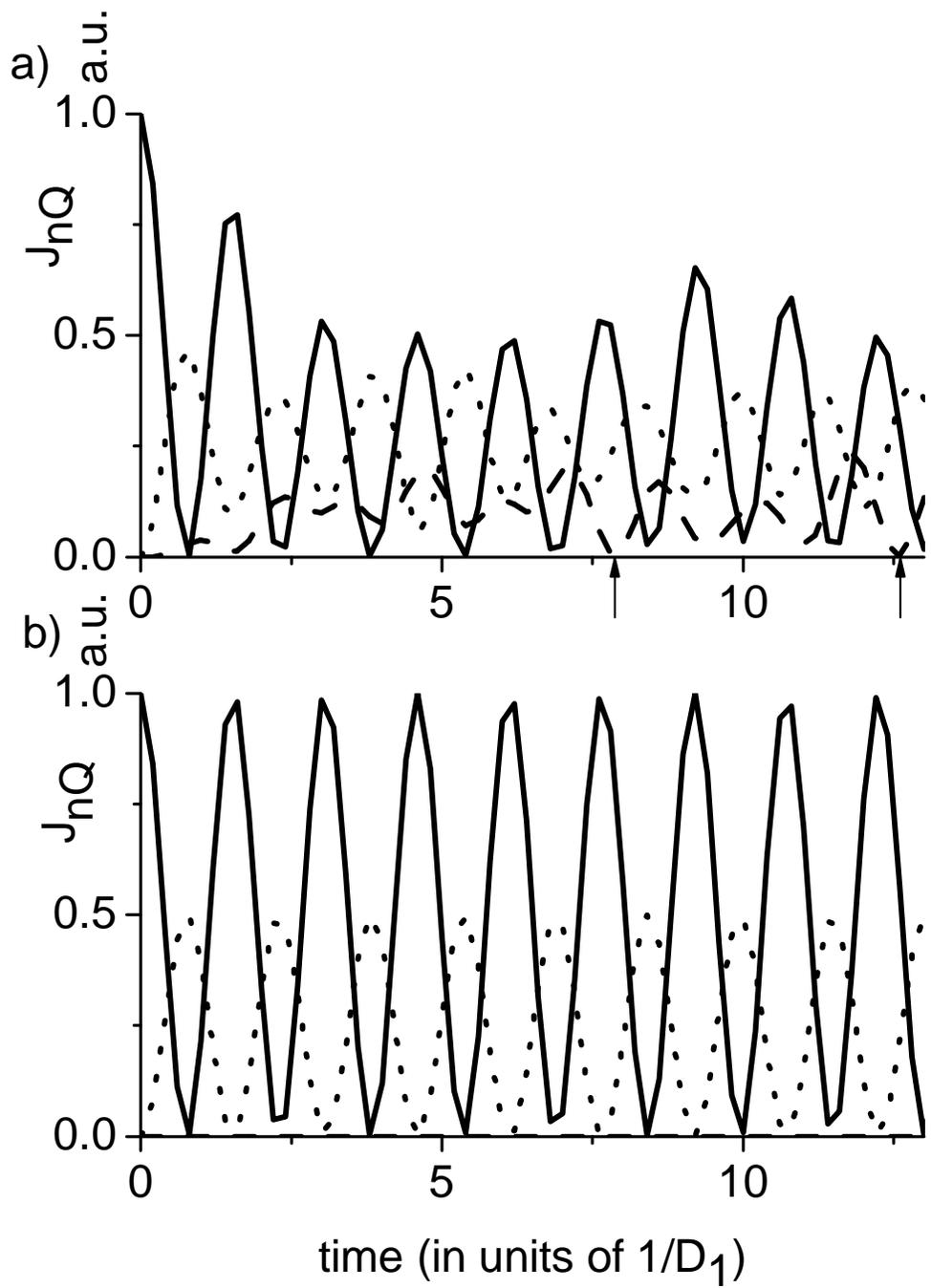

Fig.2

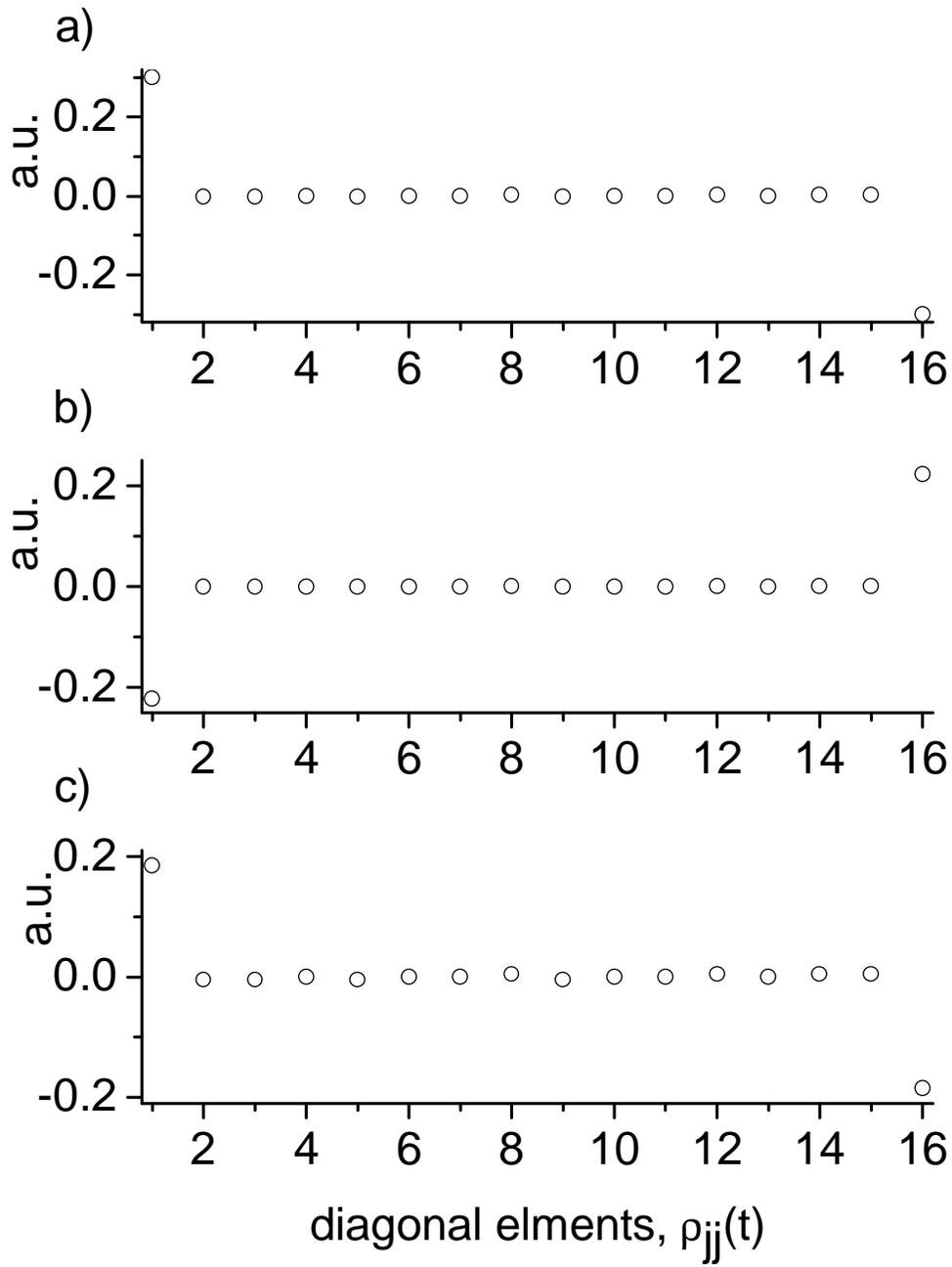

diagonal elments, $\rho_{jj}(t)$

Fig.3

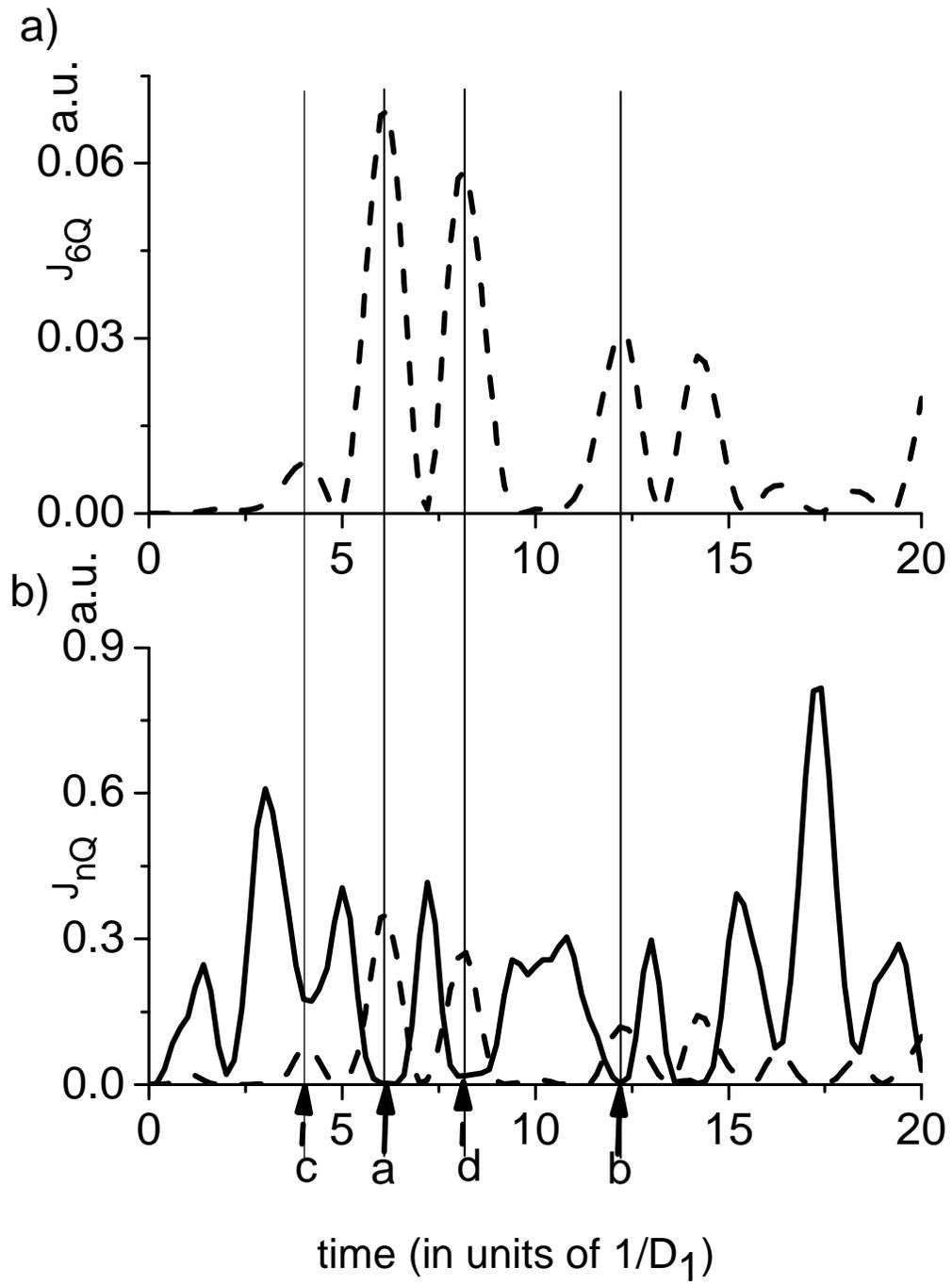

Fig. 4

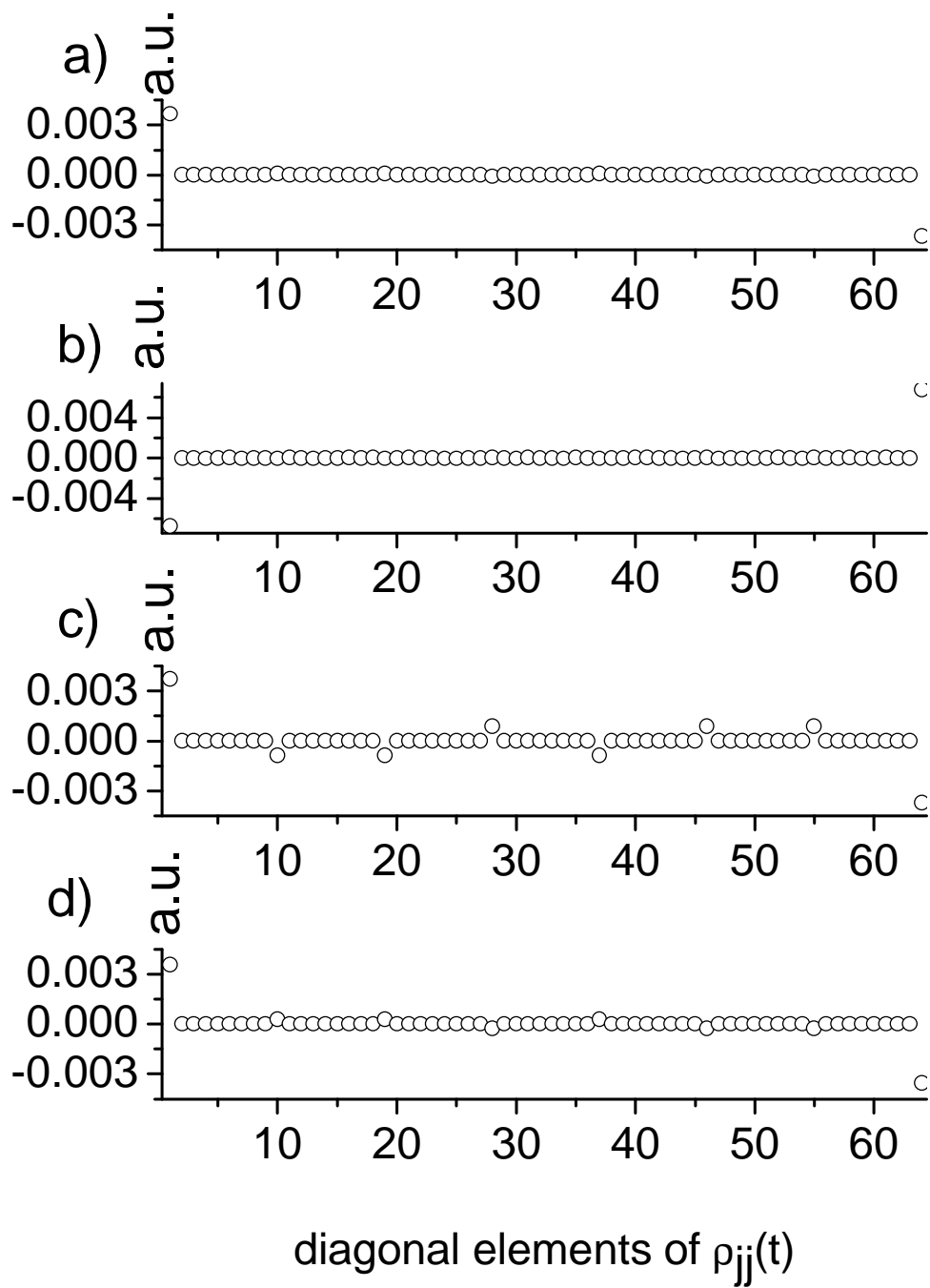

diagonal elements of $\rho_{jj}(t)$

Fig. 5

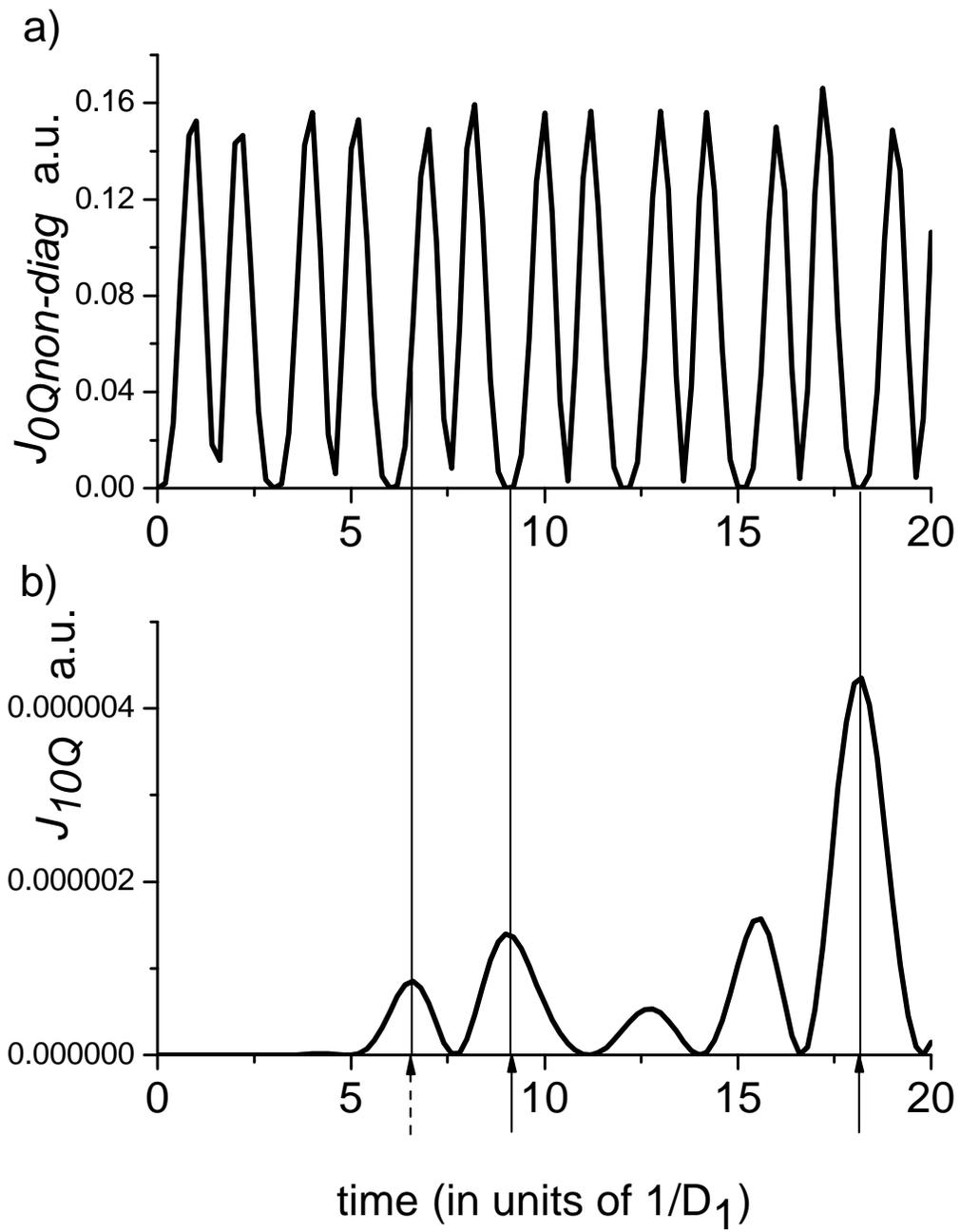

time (in units of $1/D_1$)

Fig. 6

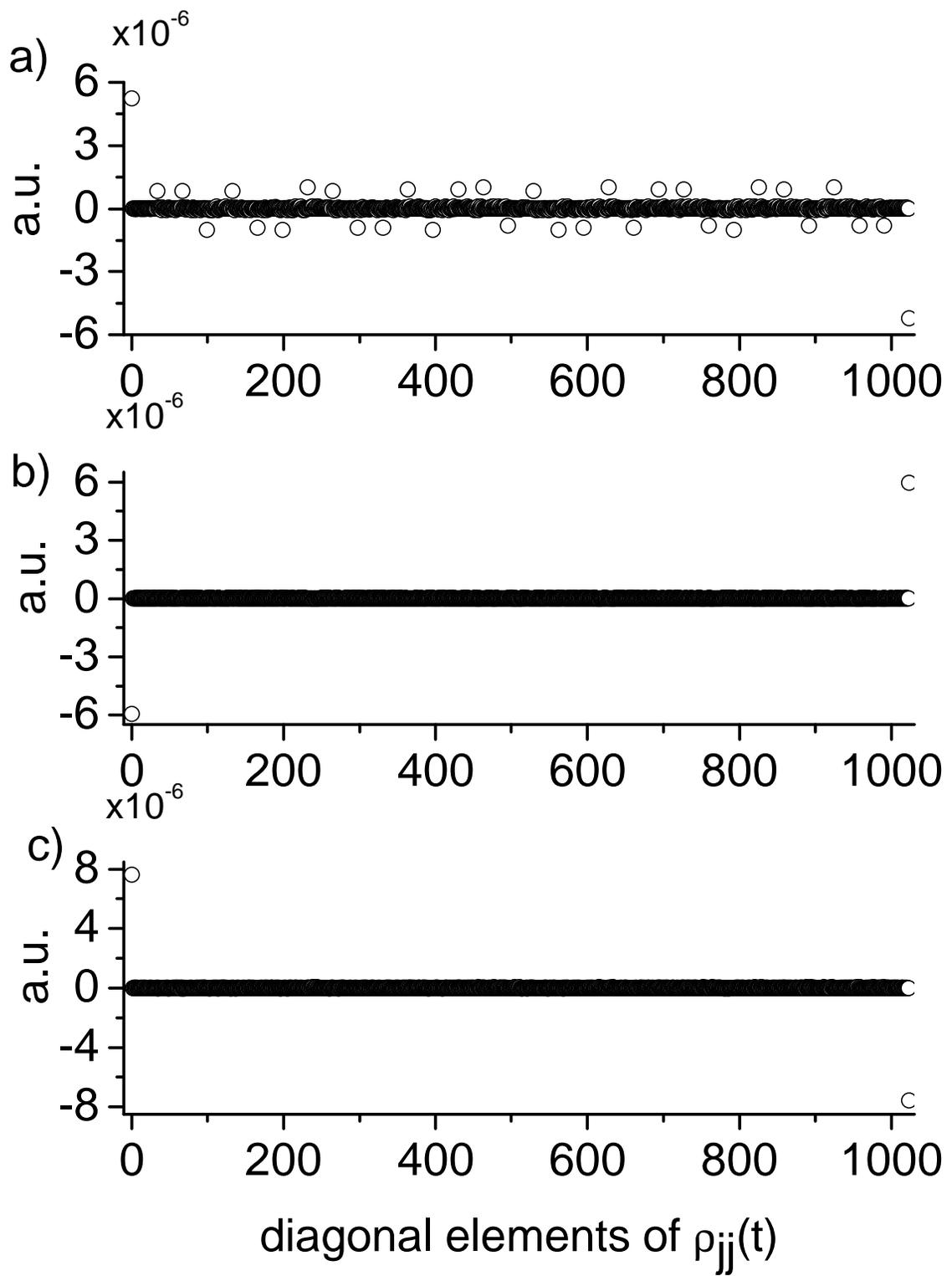

diagonal elements of $\rho_{jj}(t)$

Fig. 7